\documentclass[aps,prd,10pt,twocolumn,superscriptaddress,showpacs]{revtex4-1}
\pdfoutput=1
\usepackage{hyperref}
\usepackage{graphicx}
\usepackage{amsfonts,amsmath,amssymb,bm,bbm}
\usepackage{color}

\hypersetup{
    pdfnewwindow=true,      
    colorlinks=true,       
    linkcolor=blue,          
    citecolor=blue,        
    filecolor=blue,      
    urlcolor=blue        
}

\begin{document}

\title{\mbox{\hspace{-0.25cm} Cosmic neutrino cascades from secret neutrino interactions}}

\author{Kenny C. Y. Ng}
\email{ng.199@osu.edu}
\affiliation{Center for Cosmology and AstroParticle Physics (CCAPP), Ohio State University, Columbus, OH 43210}
\affiliation{Department of Physics, Ohio State University, Columbus, OH 43210}

\author{John F. Beacom}
\email{beacom.7@osu.edu}
\affiliation{Center for Cosmology and AstroParticle Physics (CCAPP), Ohio State University, Columbus, OH 43210}
\affiliation{Department of Physics, Ohio State University, Columbus, OH 43210}
\affiliation{Department of Astronomy, Ohio State University, Columbus, OH 43210} 

\date{August 16, 2014}

\begin{abstract}
The first detection of high-energy astrophysical neutrinos by IceCube provides new opportunities for tests of neutrino properties.  The long baseline through the Cosmic Neutrino Background~(C$\nu$B) is particularly useful for directly testing secret neutrino interactions~($\nu$SI) that would cause neutrino-neutrino elastic scattering at a larger rate than the usual weak interactions.  We show that IceCube can provide competitive sensitivity to $\nu$SI compared to other astrophysical and cosmological probes, which are complementary to laboratory tests.  We study the spectral distortions caused by $\nu$SI with a large s-channel contribution, which can lead to a dip, bump, or cutoff on an initially smooth spectrum.   Consequently, $\nu$SI may be an exotic solution for features seen in the IceCube energy spectrum. More conservatively, IceCube neutrino data could be used to set model-independent limits on $\nu$SI.  Our phenomenological estimates provide guidance for more detailed calculations, comparisons to data, and model building.
\end{abstract}


\maketitle

\section{Introduction}
\label{sec:Introduction}

Neutrinos are mysterious.  The discovery of neutrino mass and mixing established physics beyond the standard model.  With rapid improvements in experimental sensitivity, neutrinos might soon reveal more dramatic new physics. This could include signatures that depend on neutrino mass, e.g., neutrino decay, neutrino magnetic moments, or neutrinoless double beta decay.  The weak interactions of neutrinos make them unique messengers for studying astrophysical systems.  The extreme scales of astrophysics allow tests of neutrino properties far beyond what is possible in the laboratory, and may reveal new interactions that shed light on the origin of neutrino mass and other important questions.

The term ``secret neutrino interactions"~($\nu$SI) indicates new physics that couples neutrinos to neutrinos.  A wide variety of models have already been considered, and some have implications for neutrino masses.  A way to characterize these models is by their mediator mass.  For massless mediators, such as in Majoron models~\cite{Chikashige:1980ui, Gelmini:1980re, Georgi:1981pg, Gelmini:1982rr, Nussinov:1982wu}, there is at least one stable new particle.  For very heavy mediators, one can use an effective theory to study the phenomenology of a class of models~\cite{Kolb:1987qy, Bilenky:1992xn, Bilenky:1994ma, Bilenky:1999dn}.  In between, the mediator mass is more moderate, and could induce resonances~\cite{Chacko:2003dt, Chacko:2004cz, Davoudiasl:2005ks, Goldberg:2005yw, Baker:2006gm, Wang:2006jy, Gabriel:2006ns}. In some models, the neutrinos also interact with dark matter~\cite{1992hena.conf..173W, Fayet:2006sa, Barenboim:2006dj, Mangano:2006mp, Lindner:2010rr, Palomares:2011xx, Aarssen:2012fx, Shoemaker:2013tda, Dasgupta:2013zpn, Bringmann:2013vra, Farzan:2014gza}.

It is challenging to directly test $\nu$SI through neutrino-neutrino scattering.  Sufficiently high flux or volume densities of neutrinos for any interactions to occur only exist in astrophysical systems.  Even there, the difficulty is revealing (using neutrinos!)~the signatures of such interactions.  So far, only $\nu$SI interactions much stronger (in a sense explained below) than weak interactions have been constrained.  Given the difficulty of probing $\nu$SI in the laboratory, it is therefore interesting to consider more model-independent probes, such as those from astrophysics and cosmology. 

One direct probe of $\nu$SI uses astrophysical neutrinos as a beam and the Cosmic Neutrino Background (C$\nu$B) as a target.  Kolb and Turner~(hereafter KT87)~\cite{Kolb:1987qy} utilized the detection of astrophysical neutrinos from SN 1987A.  The agreement of the detected signal with the standard expectation of no neutrino scattering en route yields robust constraints on $\nu$SI.  KT87 established a phenomenological approach by considering general interactions with mediator masses either much smaller or much larger than the interaction energy.  Their constraints could be applied to many possible $\nu$SI models.

The first detection of astrophysical neutrinos since SN 1987A is the 37 events detected by IceCube~\cite{Aartsen:2013bka, Aartsen:2013jdh, Aartsen:2014gkd}.  One expects a steady stream of more events in the near future, so the precision will improve quickly.  The angular distribution of the events suggests that most, if not all, of them are extragalactic in origin.  Compared to the SN 1987A neutrinos, the IceCube neutrinos have a much longer baseline through the C$\nu$B, making them more sensitive to $\nu$SI; they have much higher energies, making them more powerful probes for massive mediators; and they have a diffuse (many-source) origin, thus averaging out the uncertainties for individual sources. 

We take advantage of this new opportunity and explore the sensitivity of IceCube to $\nu$SI.  With minimal assumptions about the interaction to reduce model dependence, we show that there are regions of parameter space where $\nu$SI could cause significant distortions to the neutrino spectrum.  Because the flux is diffuse and the shape and normalization without interactions are not known, we must look for distortions to the spectrum that have characteristic shapes.  This favors interactions with strong energy dependence, especially due to a resonance.

Our method generalizes earlier work, going beyond the pure attenuation considered in KT87 and Refs.~\cite{Keranen:1997gz, Hooper:2007jr} as well as the simplified treatment of regeneration considered in Refs.~\cite{Goldberg:2005yw, Baker:2006gm}.  We improve upon these by using the propagation equation to describe the interaction of a neutrino beam with the C$\nu$B in the presence of strong $\nu$SI.  Besides attenuation, this properly takes into account regeneration as well as multiple scattering of the parent and daughter neutrinos, i.e., a cascade.

In Sec.~\ref{sec:nusi}, we consider existing $\nu$SI constraints.  In Sec.~\ref{sec:nucnub}, we discuss the effects of $\nu$SI on astrophysical neutrino spectra.  We conclude in Sec.~\ref{sec:conclusion}.  Throughout, we use cosmological parameters for which the matter density fraction is $\Omega_{M} = 0.3$, the cosmological constant density fraction is $\Omega_{\Lambda} = 0.7$, and the Hubble function is $H(z) = H_{0} \sqrt{\Omega_{\Lambda}+\Omega_{M}(1+z)^{3}}$, where $H_{0} = 70 \; {\rm km\;s^{-1}\; Mpc^{-1}}$.

\section{Secret Neutrino Interactions}
\label{sec:nusi}

\begin{figure*}[htbp]
\resizebox{18cm}{!}{\includegraphics{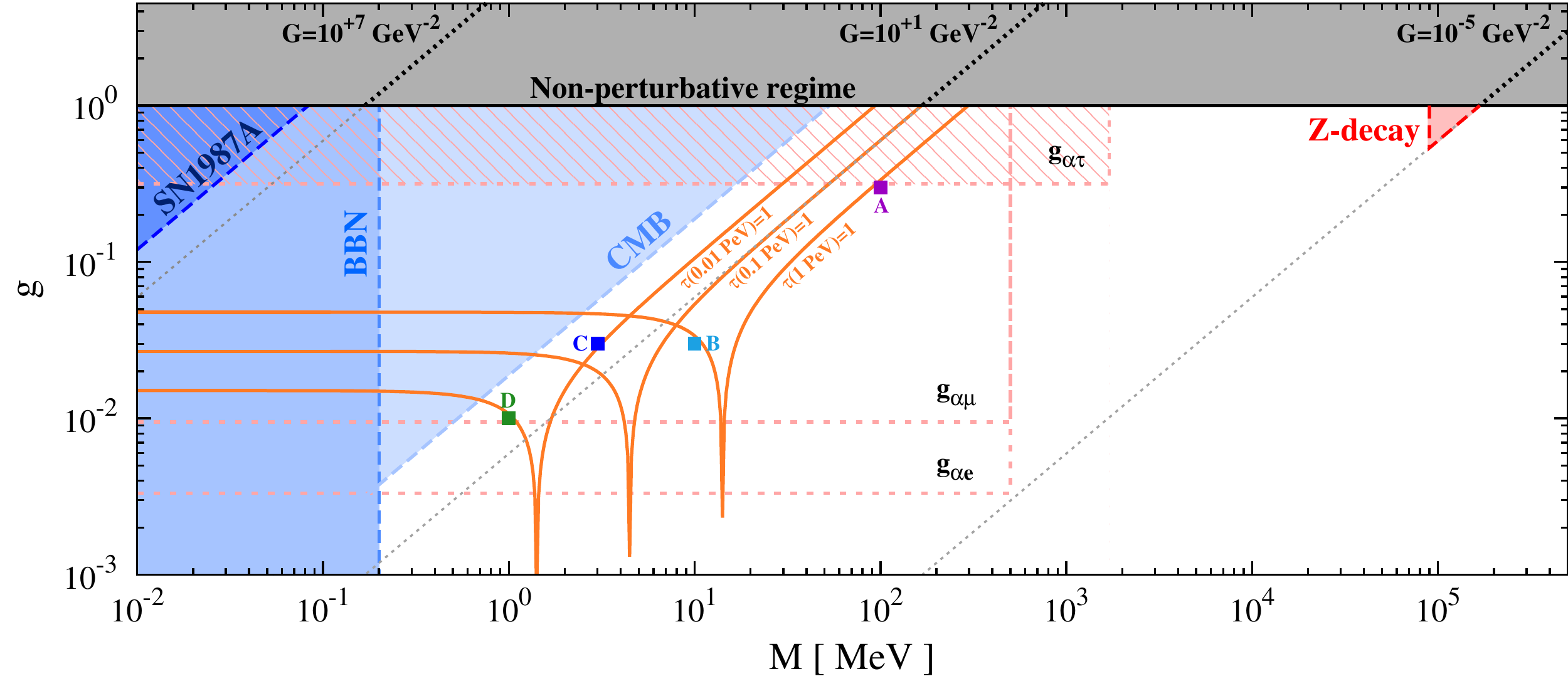}}
\caption{Present constraints and future sensitivity to $\nu$SI in terms of neutrino coupling, $g$, and mediator mass, $M$, with diagonal dotted contours shown for values of the dimensionful coupling $G$.  The blue shaded regions are excluded by astrophysical and cosmological considerations based on SN 1987A~\cite{Kolb:1987qy}, BBN~\cite{Ahlgren:2013wba}, and the CMB~\cite{Cyr-Racine:2013jua, Archidiacono:2013dua}.  The pink dashed lines indicate flavor-dependent limits based on laboratory measurements of meson and lepton decays~\cite{Lessa:2007up}; we consider only the weakest limit, for $\nu_\tau$, to be robustly excluded for all flavors, and it is shaded.  The red shaded region is excluded based on measurement of Z-boson decay~\cite{Bilenky:1999dn}.  The gray shaded region indicates the non-perturbative regime.  {\it The orange lines are contours of unit optical depth for different initial neutrino energies~({Eq.~\ref{eq:opticaldepth}}), indicating the approximate boundary of the parameter space above which IceCube is sensitive to $\nu$SI.  The squares represent the example parameters (given in Table~\ref{tab:table1}) used in our calculations. }
} 
\label{fig:constraint}
\end{figure*}

We first review existing constraints on $\nu$SI for a phenomenological scalar interaction term, ${\mathcal L}_{int}\sim g \phi \bar{\nu}\nu$.  Fig.~\ref{fig:constraint} shows the parameter space of neutrino coupling $g$ to a new mediator $\phi$ with mass $M$.  While these parameters vary depending on the specific model, type of coupling, number of new mediators, etc., this figure gives a broad comparison of different constraints.  The accuracy is adequate, especially considering the many orders of magnitude on the axes.  We do not consider the vector mediator case, since the laboratory constraints are much stronger than for the scalar case~\cite{Laha:2013xua, Karshenboim:2014tka}.

The range in mediator mass is chosen to span from the KT87 constraint to those near the weak scale, focusing on the mass range that has been of particular interest in recent work, e.g., Refs.~\cite{Aarssen:2012fx, Dasgupta:2013zpn, Bringmann:2013vra}.  Coincidentally, this is where the IceCube sensitivity is greatest, as we show below.  The range in coupling is chosen from the boundary of the non-perturbative regime to where IceCube loses sensitivity.  It would be possible to extend the figure to smaller masses and couplings, showing interesting features in some of the limits, but that would detract from our focus, so we just describe those features in the text.  

There are three kinematic regimes in which constraints can be derived, depending on how the mediator mass, $M$, compares to the interaction energy in the center of momentum frame, $\sqrt{s}$.  These are where the mediator mass is small (i.e., like a Majoron, where the mediator mass is zero or negligible), comparable to the interaction energy (i.e., where the energy dependence of the cross section depends on the mediator mass, possibly through a resonance), or large (i.e., an effective theory where the mediator mass has been integrated out).  Constraints derived assuming extremely small or large mediator masses cannot be applied beyond their domains of applicability. 

An effective theory description is appropriate when $M \gtrsim \sqrt{s}$ and $g \lesssim 1$.  Then these parameters can be characterized in a combination analogous to the Fermi constant for low-energy weak interactions, i.e., a dimensionful coupling
\begin{equation}
G \equiv \frac{1}{\sqrt{4\pi}} \frac{g^{2}}{M^{2}} \,. 
\end{equation}
Constraints on $\nu$SI are sometimes quoted using just $G$.  This does not provide as much information as a region in the $g$-$M$ plane, because of the degeneracy in $g/M$ and the unspecified limits of applicability.   In Fig.~\ref{fig:constraint}, we plot some diagonal contours of constant $G$.  The fact that the line given by $G\sim G_F \sim 10^{-5} \; {\rm GeV}^{-2}$ is not the same as the single point ($M \sim 100$ GeV, $g \sim 1$) that defines the weak interactions illustrates our caution about characterizing $\nu$SI with only $G$.

One general framework for directly testing $\nu$SI is to use neutrinos from sources that travel a long distance through the C$\nu$B. The only possibly relevant standard model process is the Z-burst scenario~\cite{Weiler:1982qy, Weiler:1997sh}, where a high energy neutrino interacts with the C$\nu$B through a Z-boson resonance.  However, the required neutrino energy is extremely high, $\sim 10^{14}\,{\rm GeV}$, and neutrinos of such energy may not exist; the cross section at lower energies is much smaller.  Any significant neutrino self-interaction observed at lower energies must be due to $\nu$SI. 

In terms of specific limits, neutrinos detected from SN 1987A were the first and, until recently, only direct detection of neutrinos from astrophysical sources beyond the Sun.  Requiring that these neutrinos travel through the C$\nu$B without scattering leads to a robust upper limit on the cross section.  Had the interaction strength been larger, the neutrinos would have scattered to lower energy and fallen below the detector sensitivity~\cite{Kolb:1987qy}.  The limit from KT87 corresponds to $G \lesssim 10^{8} \; {\rm GeV}^{-2}$.  The average supernova neutrino energy is $\sim 10$ MeV and that of the C$\nu$B is $\sim 10^{-1}\,{\rm eV}$~(assuming small but degenerate neutrino masses), making $\sqrt{s} \sim 10^{-3}$ MeV, so the applicability of this limit would end below a vertical boundary at $M \sim 10^{-3}$ MeV (not shown).  For a massless neutrino, this boundary would be at $M \sim 10^{-4}$ MeV, because the average neutrino energy is $\sim 10^{-3}\,{\rm eV}$.

Another general framework for directly testing $\nu$SI is through their effects on a gas with a high neutrino density.  In the early universe~\cite{Notzold:1987ik, Hannestad:1995rs, Dolgov:2002ab, Abazajian:2002qx, Mangano:2005cc} and in core-collapse supernovae~\cite{Qian:1994wh, Buras:2002wt, Duan:2005cp, Hannestad:2006nj, Raffelt:2007yz, Dasgupta:2009mg, Duan:2010bg, Friedland:2010sc, Cherry:2012zw, Raffelt:2013rqa}, the conditions are so extreme that even standard model scale neutrino self-scattering and their self-induced matter mixing potential are important.  In the early universe, $\nu$SI could cause neutrinos to annihilate or decay into light particles, modifying the expansion history~\cite{Masso:1994ww, Beacom:2004yd, Hannestad:2005ex, Bell:2005dr, Ahlgren:2013wba}; change the free-streaming property of neutrinos during photon decoupling~\cite{Chacko:2003dt, Hannestad:2004qu, Friedland:2007vv, Cyr-Racine:2013jua, Archidiacono:2013dua}; or induce new mixing effects~\cite{Hannestad:2013ana, Dasgupta:2013zpn}. { In supernovae, the effect of elastic scattering on neutrino escape time~\cite{Manohar:1987ec} is irrelevant~\cite{Dicus:1988jh},} but $\nu$SI could cause a phase transition~\cite{Davoudiasl:2005fd, Sher:2011mx}, change the cooling process~\cite{Choi:1989hi, Kachelriess:2000qc, Farzan:2002wx, Fayet:2006sa, Zhou:2011rc}, or induce non-standard flavor mixing \cite{Blennow:2008er}. 

There are several specific limits.  In the early universe, if the $\nu$SI mediator is not too massive, it could be in thermal equilibrium, changing the number of relativistic degrees of freedom~\cite{Masso:1994ww, Ahlgren:2013wba}, which can be tested through Big Bang Nucleosynthesis (BBN).  We show the ``maximally conservative" case from Ref.~\cite{Ahlgren:2013wba}, which assumed vector boson mediators.  The BBN limits extend down to $g\sim 10^{-8}$.  The presence of $\nu$SI can also change the free-streaming property of the C$\nu$B, which can leave an imprint on the observed Cosmic Microwave Background (CMB).  Strong constraints on $\nu$SI have been set in Refs.~\cite{Cyr-Racine:2013jua, Archidiacono:2013dua}.  In the mediator mass range we focus on, the constraint was derived assuming a heavy mediator, and is $G \lesssim 100 \;{\rm GeV}^{-2}$, which is much more stringent than the SN 1987A limit.  In Ref.~\cite{Cyr-Racine:2013jua, Archidiacono:2013dua}, the C$\nu$B is constrained to be free-streaming untill redshift $\sim 2\times10^{5}$, where $\sqrt{s}\sim 10^{-4}\,{\rm MeV}$. Therefore, the domain of applicability of the CMB limit would end at a vertical boundary~(not shown) at $M\sim10^{-4}\,{\rm MeV}$.

Limits on $\nu$SI can also be set by observations of laboratory processes.  Even if the neutrinos are not detected directly, their presence can be inferred by  precise measurements of other particles.  For example, in the presence of $\nu$SI, a mediator can be produced by bremsstrahlung from an external neutrino~\cite{Barger:1981vd, Lessa:2007up, Laha:2013xua}; a massive mediator will then decay into other particles~\cite{Bilenky:1992xn, Bilenky:1994ma, Bilenky:1999dn}.  In Majoron-type models, the best laboratory constraints come from meson and lepton decays, e.g., Refs.~\cite{Barger:1981vd, Lessa:2007up}, but they depend on the particular flavor coupling, $g_{\alpha\beta}$, where $\alpha,\beta = e,\mu,\tau$, and are valid only up to the mass of the meson or lepton, e.g., kaons or tau leptons.  The couplings involving $\nu_\tau$ are the least constrained, so we regard them as the most robust.  Accordingly, they are shaded in Fig.~\ref{fig:constraint} to indicate exclusion for all flavors.

An flavor-independent constraint can be obtained from Z-boson decay.  If a heavy mediator is assumed, the limit is quite strong, $G \sim G_{F}$, as shown in Refs.~\cite{Bilenky:1992xn, Bilenky:1994ma, Bilenky:1999dn}, though the domain of applicability of that effective theory calculation ends below the Z-boson mass.  We emphasize that though this is nominally a very strong limit, it does {\it not} rule out all of the parameter space above the diagonal line $G \sim G_{F} \sim 10^{-5} \;{\rm GeV}^{-2}$.  If the calculation is extended to allow a light mediator, following Ref.~\cite{Laha:2013xua}, the result for the scalar mediator case (not shown) is comparable to the $g_{\alpha\tau}$ constraint in Fig.~\ref{fig:constraint} for mediator masses below the mass of Z-boson.  

This combination of constraints shows a window of parameter space in the MeV range where model-independent astrophysical or cosmological constraints could be improved.  The IceCube sensitivity, shown approximately by the three orange lines in Fig.~\ref{fig:constraint} and calculated in the next section, lies in this region.  Because both the astrophysical neutrino beam and the C$\nu$B targets are expected to contain all flavors of neutrinos, the IceCube sensitivity is complementary to the flavor-dependent laboratory limits.

\section{Astrophysical Neutrino Interactions with the C$\nu$B}
\label{sec:nucnub}

\subsection{Sensitivity estimate for $\nu$SI}

We first make an order-of-magnitude estimate of the sensitivity of IceCube to $\nu$SI with cross section $\sigma_{\nu\nu}$; we present a more detailed calculation in later sections.  If the neutrino mass scale is $\simeq 0.1$ eV, target C$\nu$B neutrinos have degenerate masses and are nonrelativistic today.  Because of neutrino mixing, all flavors of neutrinos and antineutrinos should be present with comparable fractions in both the beam and target.  To be conservative, we assume that only one species (flavor or mass eigenstate) of neutrinos and antineutrinos in the beam interacts, each with only half of the targets of a given species, so the target density is $n_t \simeq 56$ cm$^{-3}$~\cite{Hannestad:2006zg, Wong:2011ip}.  A typical distance for astrophysical sources is the Hubble length, $c/H_{0} \simeq 4$ Gpc.  The optical depth for $\nu$SI interactions is $\tau \simeq n_t \sigma_{\nu\nu} c/H_{0}$.  For $\nu$SI to affect neutrino propagation appreciably, one would require $\tau \simeq 1$, and therefore 
\begin{equation}
\label{eq:sigma_0}
\sigma_{\nu\nu} \simeq
1\times 10^{-30} \, {\rm cm^{2}} \,.
\end{equation}
This is an necessary, but not sufficient, condition for the effects of $\nu$SI on astrophysical neutrinos to be observed.  The actual $\nu$SI sensitivity for an neutrino telescope depends on model details such as the resonance energy and detector details such as the analysis energy range.  A larger $\sigma_{\nu\nu}$ would severely affect the incoming neutrino beam, because the unattenuated fraction scales as $e^{-\tau}$.

If the interaction is through a heavy mediator, then $\sigma_{\nu \nu} \simeq G^2 s \simeq G^2 2 E m_\nu$.  For the PeV neutrinos detected by IceCube, this leads to a scale
\begin{equation}
\label{eq:4fermi}
G \simeq 4 \; {\rm GeV^{-2}} \,.
\end{equation}
This nominal sensitivity on G is a factor of $\sim 25$ times below than the CMB limit~\cite{Archidiacono:2013dua}.  Thus, the detection of astrophysical neutrinos gives an exciting opportunity to test $\nu$SI.

\subsection{Neutrino propagation: free streaming case}

In the usual case, neutrinos from cosmic sources travel to the detector without interaction.  The standard model neutrino self-interaction cross section grows linearly with neutrino energy when the center of mass energy is below the mass of the Z boson. It is $\sim10^{-41}\, {\rm cm^{2}}$~\cite{Roulet:1992pz, Bhattacharjee:1998qc, Barenboim:2004di, Lunardini:2013iwa} for a $1\,{\rm PeV}$ neutrino scattering on C$\nu$B~($0.1\,{\rm eV}$ masses), much smaller than the sensitivity estimated above.  Neutrino-CMB scattering is even more suppressed~\cite{Seckel:1997kk}.  Near 1 PeV, the neutrino-nucleon cross section is $\sim 10^{-33} \; {\rm cm^{2}}$~\cite{Gandhi:1995tf, Gandhi:1998ri}, but the nucleon density is only $\sim 10^{-7} \; {\rm cm^{-3}}$, due to the small baryon asymmetry, so the interaction probability is also negligible.  (For electrons as targets, both the cross section and number density are small.)  Therefore, any interactions of PeV neutrinos during propagation must be due to interactions beyond the standard model.

To describe the neutrino beam, we define the co-moving number density at a time $t$ by $n(t)$ and the differential (in energy $E$) number density by
$\tilde{n}(t,E) \equiv dn(t,E)/dE$.  The observable neutrino number flux is
\begin{eqnarray}
J(E) & \equiv & \frac{dN_{\nu}}{dA dt d\Omega dE} \\
& = & \frac{c}{4 \pi} \tilde{n}(0,E) \nonumber \,.
\end{eqnarray}
The evolution of the number density is described by the propagation equation~\cite{Lee:1996fp, Bhattacharjee:1998qc, Berezinsky:2002nc, Ahlers:2009rf, Ahlers:2010fw,Murase:2012xs}.  In the free-streaming limit, that is
\begin{equation}
\label{eq:freestreaming}
\frac{\partial \tilde{n}(t,E)}{\partial t} = \frac{\partial}{\partial E}(b \tilde{n}(t,E) ) + {\cal L}(t,E)\, .
\end{equation}
The first term on the right takes into account the continuous energy loss rate $b=H(t)E$ due to redshift, and the second is the differential number luminosity density of the sources.  Throughout this work, we solve the propagation equation numerically in redshift variables, conservatively taking the initial condition $\tilde{n}(z_{\rm max} = 4,E) = 0$.

In the free-streaming case, $J(E)$ has a convenient closed form in the redshift variable~\cite{Ahlers:2009rf, Berezinsky:2002nc}, given by
\begin{equation}
\label{eq:freestreaming2}
J(E) =  \int_{0}^{z_{\rm max}} \frac{c dz}{4 \pi H(z)} {\cal L}(z,E(1+z)) \, ,
\end{equation}
following from the simple relationship between emitted and observed energy as a function of redshift.  For the source term, we assume a universal emission spectrum with a factorized form, 
\begin{equation}
{{\cal L}(z,E)} = {\cal W}(z) {\cal L}_{0}(E)\,,
\end{equation}
where ${\cal L}_0(E)$ is the differential number luminosity for each source and ${\cal W}(z)$ the redshift evolution of the source density, assumed to follow the star formation rate (SFR)~\cite{Hopkins:2006bw, Yuksel:2008cu}.  The term $H(z)$ increasingly suppresses the importance of flux contributions from higher redshifts. The SFR evolution, which rises by an order of magnitude between $z = 0$ and $z = 1$ and then begins to fall, overcomes this effect until $z = 1$, so $z = 1$ is the typical redshift of the most relevant cosmic sources.

\subsection{Neutrino propagation: $\nu$SI case}

When neutrinos interact with the C$\nu$B, the effects can be calculated by adding terms to the propagation equation, so that
\begin{eqnarray}
\label{eq:boltzmann}
\frac{\partial \tilde{n}(t,E)}{\partial t}
& = & \frac{\partial}{\partial E}(b \tilde{n}(t,E) ) + {\cal L}(t,E) \\ 
& - & c n_{t} \sigma \tilde{n}(t,E)
+ c n_{t} \int_{E}^{\infty} \! dE^{\prime} \, \tilde{n}(t,E^{\prime}) \sum_{i}\frac{d\sigma_{i}}{dE} \,. \nonumber
\end{eqnarray}
We assume that the neutrino (or other particle) targets are non-relativistic; if they are not, their energy distribution needs to be taken into account in the propagation equation.  The third term on the right accounts for attenuation at a given energy due to scattering with cross section $\sigma(E)$.  The fourth term accounts for particle regeneration from one energy to another, including when an incoming particle of energy $E^\prime$ is down-scattered to a lower energy $E$ but not lost and when a target particle with their rest mass energy is up-scattered to energy $E$ to join the beam.  The distributions of secondary particles are described by the differential cross sections $d\sigma_{i}(E,E^\prime)/dE$, where $i$ denotes each  process.  Here, we only include down-scattering and up-scattering with neutrino targets; this could be generalized.  The net effect is therefore a distortion of the beam spectrum in a way that conserves energy but not particle number.

This propagation equation automatically takes into account the re-scattering of secondary particles, analogous to electromagnetic cascades for high energy cosmic gamma rays~\cite{Lee:1996fp, Murase:2012xs}, $\nu_{\tau}$ regeneration in matter~\cite{Halzen:1998be, Dutta:2000jv, Beacom:2001xn}, and ultrahigh energy cosmic ray propagation~\cite{Bhattacharjee:1998qc, Berezinsky:2002nc, Ahlers:2009rf, Ahlers:2010fw}.  As far as we know, cascade calculations have not been done for neutrino-neutrino interactions.  (A similar formalism for supernova neutrinos interacting with dark matter appeared in preprint while we were finalizing this work~\cite{Farzan:2014gza}.)

We assume that there are only active neutrinos in the beam and target, that all species are comparably populated by mixing, and that this happens long before any effects due to propagation.  We assume the neutrino masses are all $\simeq0.1$\,eV and that only one species of $\nu + \bar{\nu}$ interacts, each with half of the targets of a given species, so $n_{t}(z) = 56 (1+z)^{3} \, {\rm cm^{-3}}$  We ignore the possibility of flavor changes in scattering.  We take a generic form for the total and differential cross sections to minimize model dependence.  The propagation equation and our calculations could be generalized to account for changes in the assumptions, and we discuss below what happens when some of them are relaxed.

We focus on elastic scattering of neutrinos in the $s$ channel.  For generality, we assume that the cross section takes a Breit-Wigner form,
\begin{equation}
\label{eq:Breit}
\sigma(E) = \frac{g^{4}}{4 \pi} \frac{s}{(s-M^{2})^{2}+M^{2}\Gamma^{2}} \,,
\end{equation}
where $s = 2 E m_\nu$ and the decay width of the mediator is $\Gamma = g^{2} M / 4\pi$.  With this form, we generalize the phenomenological approach of KT87 to include the possibility that a resonance dominates the cross section.  In that case, the $t$ channel contribution can be neglected.  In the off-resonance case, neglecting the $t$ channel does not change the results much for the cases considered here.  For the differential cross section, we take a flat distribution in $E$ between zero and $E^{\prime}$, which corresponds to the case of a scalar mediator.  Vector mediators have a different distribution, but we do not consider this case due to strong constraints~\cite{Laha:2013xua, Karshenboim:2014tka}.

This form of the cross section parametrizes all three kinematic regimes of how the mediator mass compares to the interaction energy in the center of momentum frame.  For a very light mediator, the cross section is decreasing with neutrino energy, $\sigma \simeq g^{4}/(4\pi s)$, while for a very heavy mediator, the cross section is increasing with neutrino energy, $\sigma \simeq g^{4}s/(4\pi M^{4})$.  These two limits correspond to the massless and massive mediator cases considered in KT87.  For the former, the cross section is independent of $M$; for the latter, it is degenerate in $g/M$.  Between these two limits, the cross section is peaked at the resonance energy defined by $s = 2 E_{\rm res} m_{\nu} = M^{2}$, where the cross section is regulated by its decay width and is $\sigma = 4\pi/M^{2}$.  

Figure~\ref{fig:constraint} shows all three of these behaviors in the optical depth for neutrino scattering.  We define this as 
\begin{equation}
\label{eq:opticaldepth}
\tau(E| g,M) = c \int_{0}^{1}dz \frac{n_{t}(0)}{H(z)} (1+z)^{2} \sigma(E|g,M)\,,
\end{equation}
where $z = 1$ is a typical redshift for cosmic sources.  For simplicity, we ignore the redshift dependence of $\sigma(E)$ inside the integral, which would slightly broaden the range of $M$ for which a resonance could occur.  The factor $(1+z)^{2}$ comes from the target density evolution factor $(1+z)^{3}$ and a factor of $(1+z)^{-1}$ from $|dt/dz|$.  Taking redshift into account only increases $\tau$ by about a factor of 3; it would matter more if high-redshift sources were dominant.  We show contours of $\tau = 1$ for $E=$ 0.01, 0.1 and 1 PeV in Fig.~\ref{fig:constraint}.  Above the contours ($\tau>1$), the effect of scattering increases exponentially with $\tau$, which increases as $g^{4}$ for fixed $M$.  Near the sharp dips, the realistic sensitivity would be reduced by the effects of detector energy resolution.  We emphasize that we use $\tau$ just for illustration; for our main result, we calculate spectra using Eq.~(\ref{eq:boltzmann}). 


\subsection{Line emission with $\nu$SI}
\label{sec:Line}

Before considering astrophysical scenarios with broad energy spectra, it is instructive to show the effects of $\nu$SI on a mono-energetic neutrino spectrum.  For rough consistency with the IceCube data, we choose 1~PeV for the line energy and define the flux in the same units as the deduced IceCube spectrum.

\begin{figure}[t]
\includegraphics[angle=0.0, width=8.1cm]{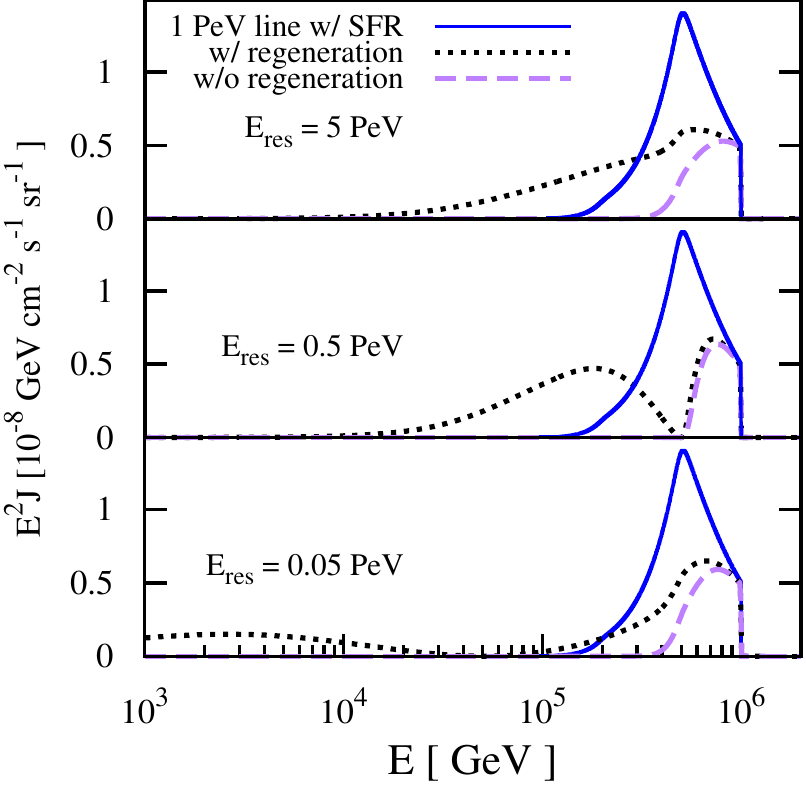}
\caption{The effects of $\nu$SI on an emitted line spectrum at 1~PeV with SFR evolution, for different assumed resonance energies, as labeled.  The solid lines are for free-streaming neutrinos, the dotted lines are for
$\nu$SI with regeneration of two neutrinos (the beam and target), and the dashed lines are for $\nu$SI with no regeneration.}
\label{fig:astro_line}
\end{figure}

In Fig.~\ref{fig:astro_line}, we compare cases with free streaming, $\nu$SI with attenuation only, and $\nu$SI with attenuation and regeneration.  The energy dependence of the $\nu$SI effects depends on the neutrino-neutrino cross section.  Using the general form of Eq.~(\ref{eq:Breit}), we choose example resonance energies $E_{\rm res}$ well above, near, and well below the emission energy of 1 PeV.  For each $E_{\rm res}$, the couplings are tuned so that $\sim e^{-1}$ of the energy spectrum is unattenuated after propagation.

For the free-streaming case, as in Eq.~(\ref{eq:freestreaming2}), the spectrum of neutrino energies simply reflects the distribution of source redshifts through the relation $E = 1\,{\rm\ PeV} / (1 + z)$.  The edge at $1\,{\rm PeV}$ is from emission nearby, the peak near $0.5\,{\rm PeV}$ from emission near $z = 1$, and lower energies from emission at still-higher redshifts.  As noted, the rapid rise of the SFR between $z = 0$ and $z = 1$~\cite{Hopkins:2006bw, Yuksel:2008cu} overcomes the suppression due to the Hubble expansion, i.e., less volume per redshift interval, until near $z = 1$, where the SFR begins to flatten and then fall.

\begin{figure}[t]
\includegraphics[angle=0.0, width=8.5cm]{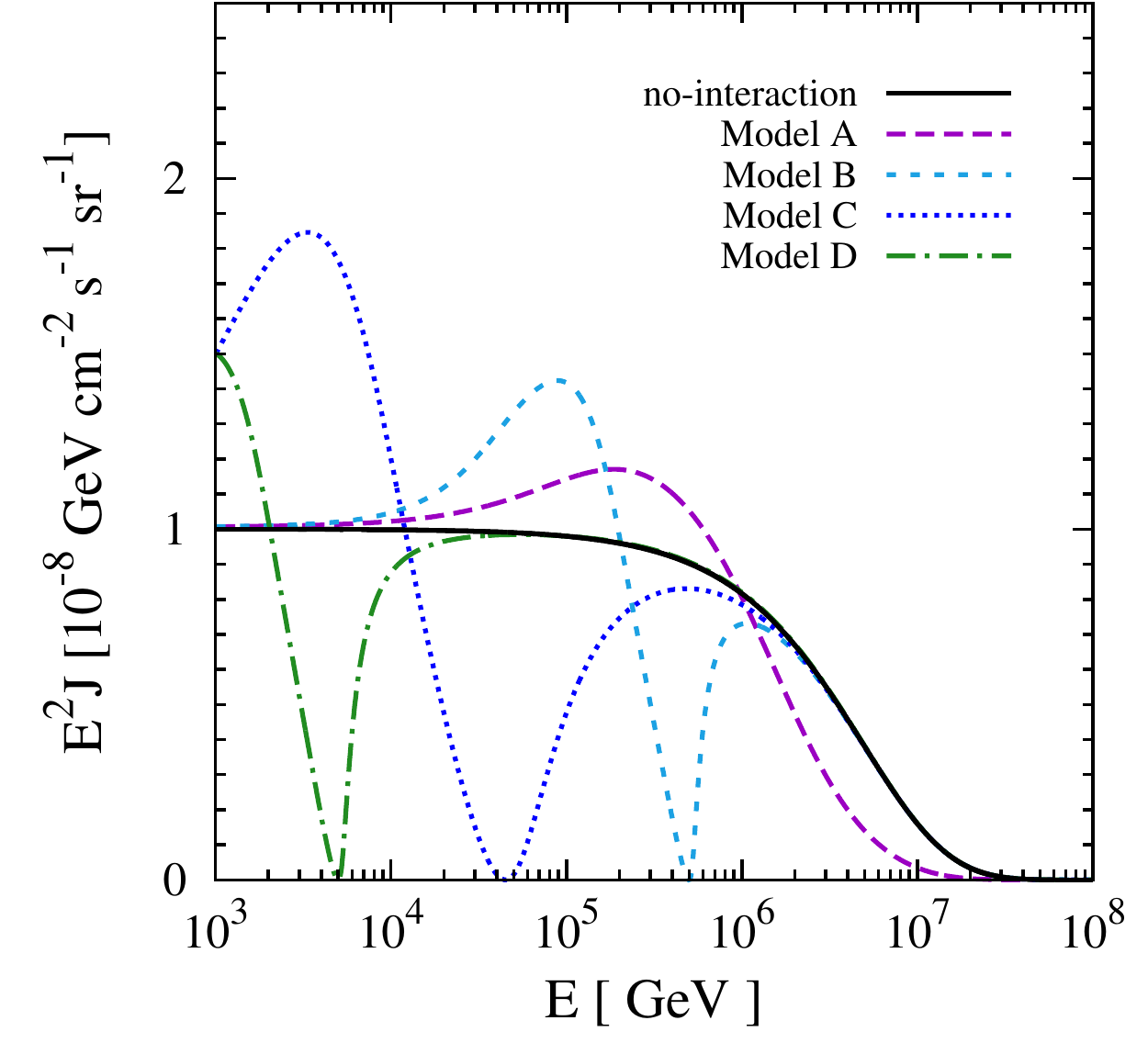}
\caption{The effects of $\nu$SI on an emitted continuum spectrum that is consistent with IceCube data.  The solid line indicates the free-streaming case ($\gamma = 2$ and $E_{\rm cut} = 10^{7}\,{\rm GeV}$), while the other lines are for four models of $\nu$SI, with the parameters defined in Table~\ref{tab:table1}.}
\label{fig:astro1}
\end{figure}

With $\nu$SI, interpreting the spectrum shape is more complicated.  Attenuation is energy-dependent, and regeneration moves particles to lower energies while increasing their numbers.  We obtain the resultant spectra by numerically solving the propagation equation, Eq.~(\ref{eq:boltzmann}).  We show the effects of attenuation alone as a comparison for the full calculation that includes regeneration.  For regeneration, the total energy carried in neutrinos is conserved.  We checked our numerical results by comparing the total energy in the spectrum to that of the free-streaming case, finding agreement at the percent level.  Energy conservation corresponds to area conservation in a plot of $E^{2} dN/dE$ with log energy bins.  In Fig.~\ref{fig:astro_line}, the area conservation is apparent, with the exception of the bottom panel, where we cut off the figure before showing the whole regenerated spectrum.

For the top panel, where $E_{\rm res} = 5\,{\rm\ PeV}$, the cross section increases with energy, which produces a flattish spectrum of regenerated neutrinos.  For the middle panel, where $E_{\rm res} = 0.5\,{\rm\ PeV}$, the resonance energy is at the peak of the unattenuated spectrum, which causes significant absorption there and a pileup of regenerated events at slightly lower energy.  Importantly, the absorption dip is broadened by the redshift effects.  Neutrinos emitted with the same energy at different redshifts reach the resonance energy at different redshifts, smearing out the resonance in received energy~(this is analogous to the redshift smearing of a monoenergetic emission line).  This broadening is helpful for detection, as a narrow feature would be difficult to observe with realistic detector energy resolution.  For the lower panel, where $E_{\rm res} = 0.05 {\rm\ PeV}$, the cross section is decreasing at energy higher than the resonance, so regenerated neutrinos will continue to interact until they fall below the resonance energy, where the energy dependence changes, forming a true cascade.

\begin{table}[b]
\caption{\label{tab:table1}Parameters for the benchmark scenarios.}
\begin{ruledtabular}
\begin{tabular}{lcccccc}
                                                                                & A                                     & B                     & C                     & D                                                    \\[0.5mm]
\hline
$g$                                         		                      & $0.3$    	                    	     & $0.03$                        & $0.03$               	        & $0.01$              \\[0.5mm]
$M \left[ {\rm MeV} \right]$                      	              & $100$    		                     & $10$                           & $3$                    	 	& $1$                   \\[0.5mm]
$\tau\left( 1\;{\rm PeV}\right )$                               & $\sim 0.7$         	             & $\sim 0.6$                   & $\sim 0.2$           		& $\sim 0.002$     \\[0.5mm]
$E_{\rm res} \left[ {\rm PeV}\right ]$                       & $50$                                   & $0.5$                          & $0.045$                          & $0.005$             \\
\end{tabular}
\end{ruledtabular}
\end{table}


\begin{figure}[t]
\includegraphics[angle=0.0, width=8.5cm]{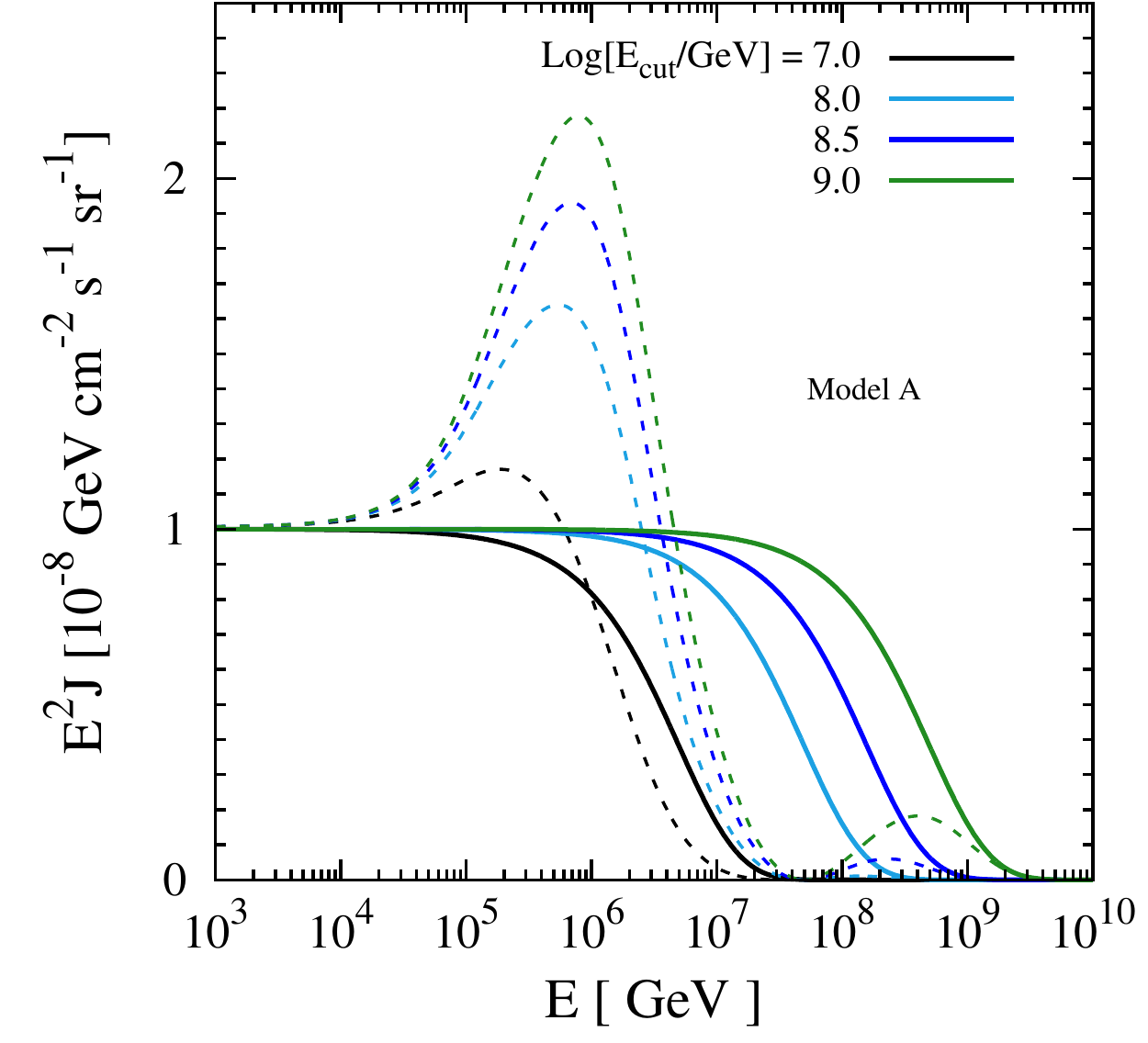}
\caption{Same as Fig.~\ref{fig:astro1}, but with different values of $E_{\rm cut}$ for the emitted spectrum.  The solid lines are the free streaming case ($\gamma = 2$ and values of $E_{\rm cut}$ as labeled), and the dashed lines are for Model A $\nu$SI.}
\label{fig:astro2}
\end{figure}


\subsection{Astrophysical scenarios with $\nu$SI}
\label{sec:Astro}

We now consider more realistic astrophysical scenarios that are compatible with IceCube measurements~\cite{Aartsen:2013bka, Aartsen:2013jdh, Aartsen:2014gkd, Laha:2013lka, Chen:2013dza, Anchordoqui:2013dnh}.  We assume that a generic astrophysical flux can be described by ${\cal L}_{0}(E) \propto E^{-\gamma}e^{-E/E_{\rm cut}}$.  An $E^{-2}$ power-law is a typical astrophysical neutrino spectrum.  IceCube detected no events above about 2~PeV.  A cutoff in the spectrum is required to explain this, because the effective area is rising, especially due to the Glashow resonance.  We first choose $E_{\rm cut} = 10^{7}\,{\rm GeV}$.  We normalize the spectrum to about the level seen by IceCube, $E^2 J(E) \sim 10^{-8}\, {\rm GeV\, cm^{-2}\, s^{-1}\, sr^{-1}}$ for neutrinos plus antineutrinos for each flavor.  Not only is this consistent with IceCube data, it is also predicted from many astrophysical scenarios~\cite{Anchordoqui:2009nf}. 

In Table~\ref{tab:table1}, we provide details for the four example points identified in Fig.~\ref{fig:constraint}.  These points are chosen to represent the regions of parameter space where $\nu$SI can have appreciable effects on the IceCube data while being consistent with the most robust limits.

In Fig.~\ref{fig:astro1}, we show the results for a continuum injection spectrum, for the free-streaming case and the four benchmark $\nu$SI cases including regeneration.  For Models B, C, and D, the presence of a resonance causes a dip and a pileup of the cascaded neutrinos right below the dip.  For Model A, the spectrum cutoff is steepened and a small pileup is produced.  These features are potentially large enough to be observed, and may even explain the gap seen in the IceCube spectrum at moderate energies.  

The lowest energy events currently observed by IceCube are $\simeq 0.03\,{\rm PeV}$, which means small mediator masses like those for Model D are difficult to observe through an obvious dip.  Of course, even if the resonance energy is below the detector energy threshold, its effects can be observed if the coupling is large enough.  At the other extreme, the largest mediator masses that can be observed through an obvious dip depend on the highest observed neutrino energies.  Similarly, even larger mediator masses can be probed if the coupling is large enough.

In Fig.~\ref{fig:astro2}, we show the effects of extending the spectrum to higher energies with and without $\nu$SI.  Without $\nu$SI, these spectra extend to energies well above what IceCube has observed, and therefore are unrealistic.  However, it is possible that the emitted spectrum does extend to high energies, but $\nu$SI lead to the observed cutoff near 1 PeV.  We use Model A as an example, which has a high mediator mass and a resonance energy of $50$ PeV.  An accompanying cascade bump occurs at $\sim1\,{\rm PeV}$, with the height of the bump reflecting the energy carried by higher-energy neutrinos that down-scattered.  

It might be possible that IceCube is not seeing a dip below PeV energies, but rather a bump near 1~PeV.   This would require a higher $\nu$SI energy cutoff and therefore a larger mediator mass than that of Model A.  This scenario, however, is in tension with other constraints because a larger coupling is needed for a larger mediator mass to maintain the same interaction strength.  Nonetheless, this could be an interesting scenario~\cite{Ioka:2014kca}, and we discuss the impact on the measured event spectrum more in the next subsection. 

\begin{figure}[t]
\includegraphics[angle=0.0, width=8.5cm]{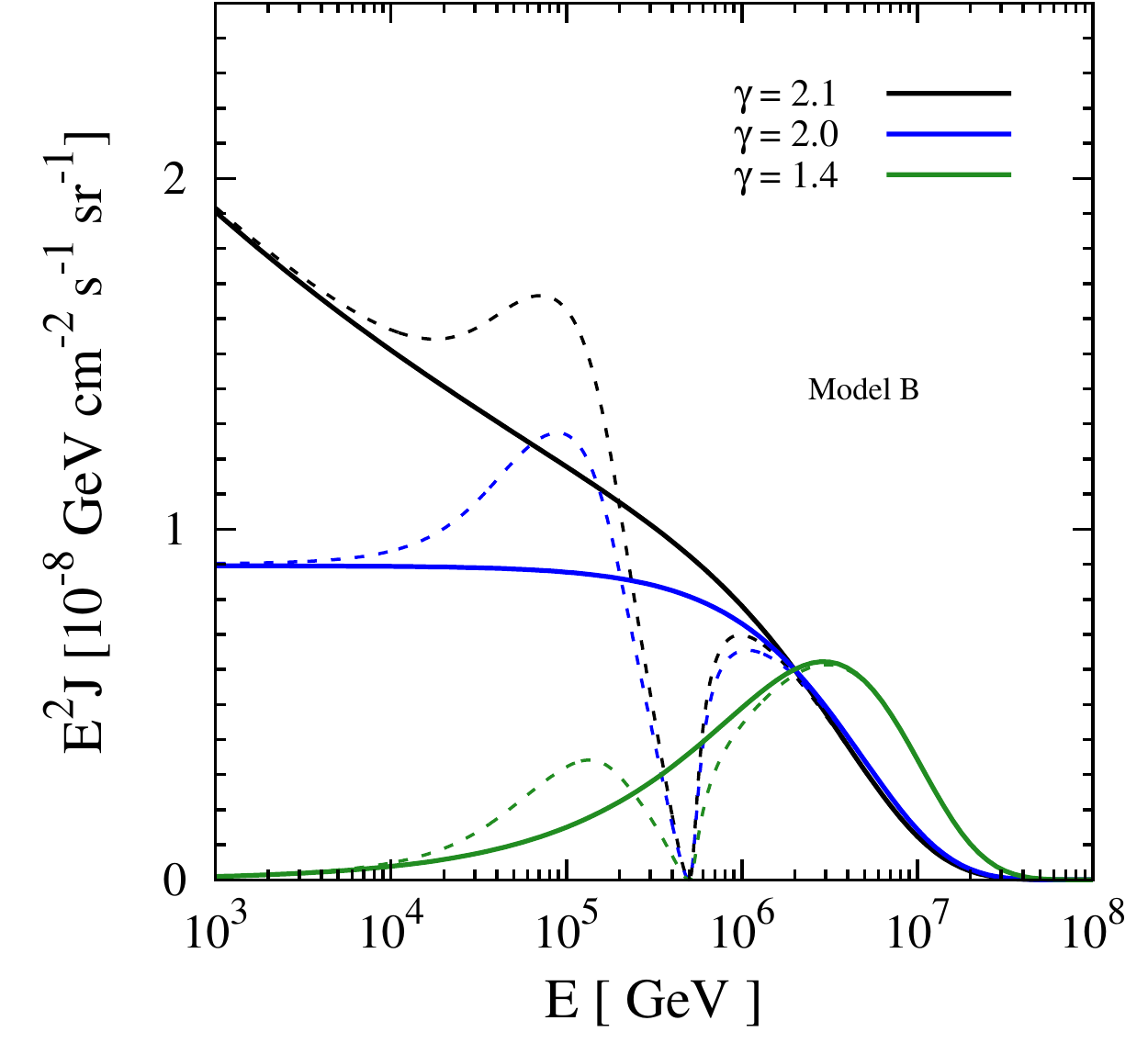}
\caption{Same as Fig.~\ref{fig:astro1}, but with different values of the spectral index.  The solid lines are for the free-streaming cases (values of $\gamma$ as labeled and $E_{\rm cut} = 10^{7}\,{\rm GeV}$), and the dashed lines are for Model B $\nu$SI.}
\label{fig:astro3}
\end{figure}

In Fig.~\ref{fig:astro3}, we show the effects of changing the spectral index.  We use Model B as an example, which has its resonance energy slightly below the spectrum cutoff.  We normalize the spectra to be the same at $2\,{\rm PeV}$.  The cascade bump makes the spectrum harder below the absorption dip.  The $\gamma = 1.4$ case roughly mimics an astrophysical spectrum with a $p\gamma$ origin~(e.g.,~\cite{Takami:2007pp, Murase:2013ffa}), as opposed to a flatter power-law spectrum with a $pp$ origin~(e.g.,~\cite{Murase:2013rfa}).  For this case, the spectrum with $\nu$SI can have twin bumps, separated by an absorption dip.


\subsection{Detection prospects}
Here we assess the prospects for detecting distorted neutrino spectra in IceCube.  We focus on cascade events, because they have a large signal to background ratio and because they reflect the underlying neutrino energy spectrum better than track events~\cite{Beacom:2004jb}.  Cascade events with energy $\lesssim$\,1\,PeV are caused by $\nu_{e}$ and $\nu_{\tau}$ charged-current reactions, plus a small contribution from all-flavor neutral-current reactions, while $\nu_{\mu}$ charged-current reactions cause only track events.  We follow~\cite{Laha:2013lka} and compute the cascade energies deposited in the detector, taking into account the different mean cascade energy for different interaction modes.  

\begin{figure}[t]
\includegraphics[angle=0.0, width=8.5cm]{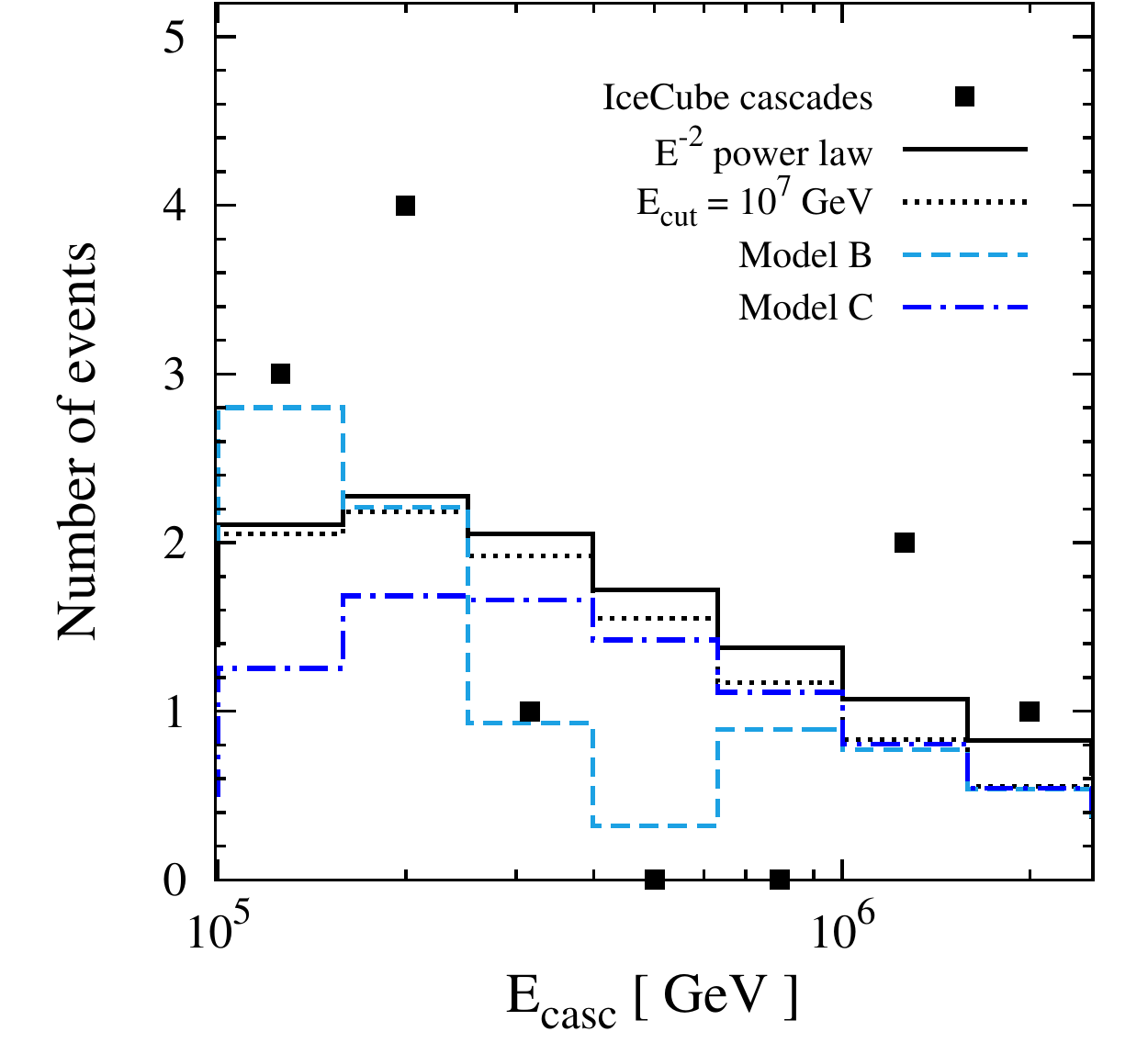}
\caption{The predicted spectra of cascade events\,(histogram), taking into account realistic detector effects, compared to the 988 days of existing IceCube data\,(points).  Selected curves from Fig.~\ref{fig:astro1} are shown, along with an $E^{-2}$ power law with no intrinsic cutoff. }
\label{fig:binspec}
\end{figure}

We comment on one possible flavor scenario using the cases depicted in Fig.~\ref{fig:astro1} as an example. We assume the initial flavor ratio to be $\nu_{e}$:$\nu_{\mu}$:$\nu_{\tau}$\,=\,1:2:0 for both neutrino and anti-neutrinos, and we take the only non-zero $\nu$SI coupling to be $g_{\tau\tau}$, in light of strong constraints on other flavors.  The $\nu$SI mean free path is much longer than the neutrino oscillation lengths, so it is safe to assume that neutrinos propagate as an incoherent mixture of mass states with ratio $\nu_{1}$:$\nu_{2}$:$\nu_{3} \simeq$\,1:1:1 before they interact with the $\nu_{\tau}$ content of the C$\nu$B.  For an imagined case where every mass state was 1/3 $\nu_{\tau}$, then each state would be depleted equally, though with 1/3 of the interaction strength compared to cases shown above, where we considered just the one flavor scenario. 

Realistically, because the $\nu_{\tau}$ fraction is non-negligible but different in each of the three mass states, each interacts with the C$\nu$B with the modified cross section $|U_{i\tau}|^{2}\sigma_{\nu\nu}$, where $U_{i\alpha}$ is the standard neutrino mixing matrix.  The two up- and down-scattered $\nu_{\tau}$ states rapidly mix to the mass state ratio $\sim$\,0.3:1.2:1.5, which will appear in the detector with flavor ratios $\sim$\,0.6:1.1:1.3.  The reduction of $\nu_{e}$ cascade events is mostly compensated by the increase in $\nu_{\tau}$ cascade events.  We neglect the accumulated flavor effects for multiple regenerations, and assume the final flavor ratio remains $\sim$\,1:1:1.  Finally, in order for each mass state to have the spectrum shown in Fig.~\ref{fig:astro1}, we would need to compensate the factor $|U_{i\tau}|^{2}$ by increasing $\sigma_{\nu\nu}$.  The smallest element is $|U_{1\tau}|^{2}\sim0.1$, therefore \emph{at most} it suffices to increase $g$ by $\sim1.8$ for all flavor spectra to have \emph{at least} the same degree of spectral distortion as in Fig.~\ref{fig:astro1}.  Considering the constraints on our benchmark models, this would be viable for all but Model A. 


\begin{figure}[t]
\includegraphics[angle=0.0, width=8.5cm]{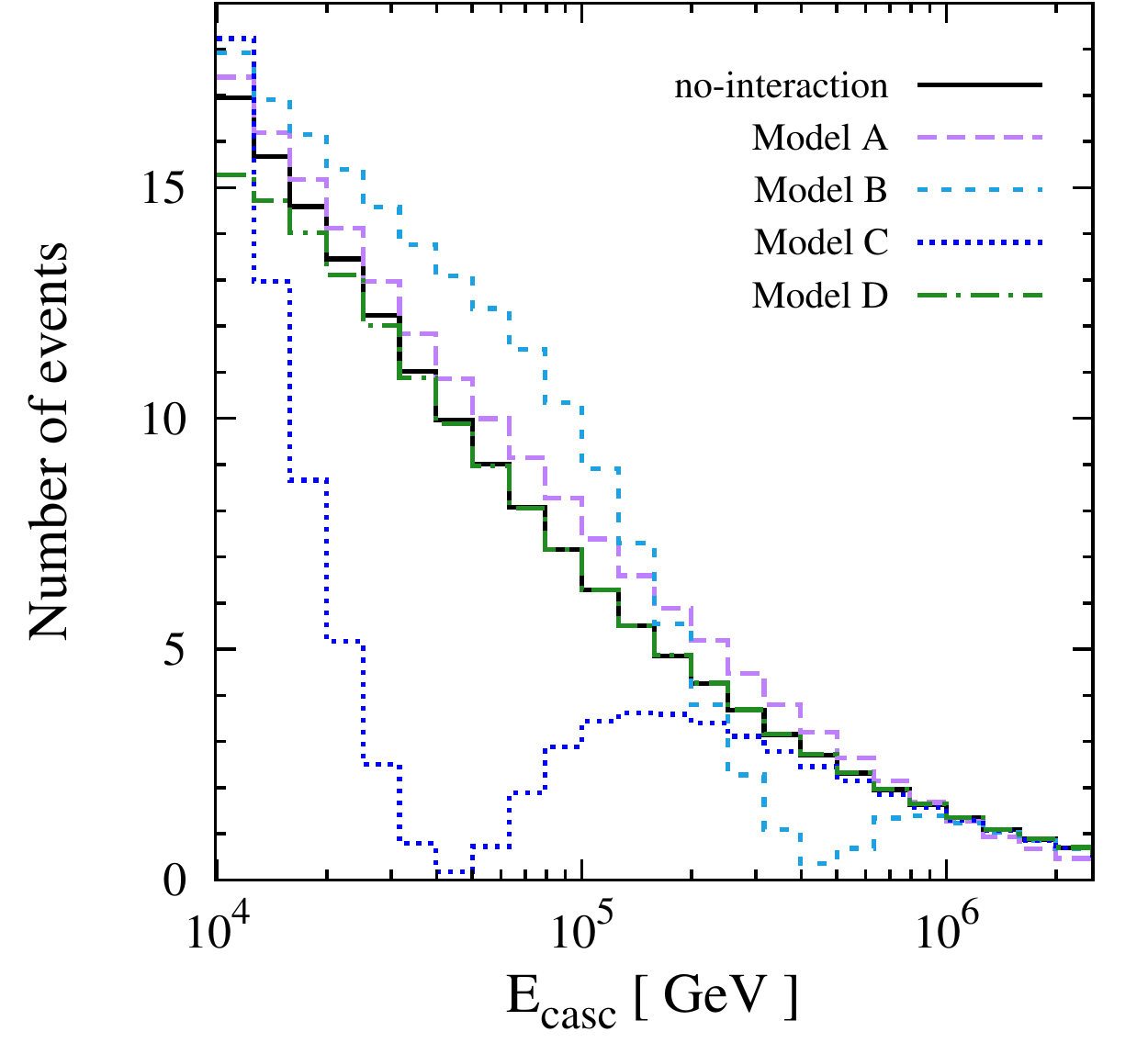}
\caption{The same as Fig.~\ref{fig:binspec}, except that the calculations are for 988 days of an ideal efficiency 1-Gton detector, the energy bins are half as wide, and the range extends to lower energies. }
\label{fig:ednde}
\end{figure}

In Fig.~\ref{fig:binspec}, we show the binned cascade event spectrum detected by IceCube.  We consider only events with deposited energy above $10^{5}\,{\rm GeV}$, because below that energy the background events shown in the IceCube paper are comparable to the signal.  For comparison, we show the expected number of events from an unbroken $E^{-2}$ power law, the same with $E_{cut} = 10^{7}\,{\rm GeV}$, and the $\nu$SI spectra for Models B and C from Fig.~\ref{fig:astro1}.  We assume $E^{2}J(E) = 10^{-8}\, {\rm GeV\,cm^{-2}\,s^{-1}\,sr^{-1}}$ below the cutoff, which is consistent with the IceCube data~\cite{Aartsen:2014gkd}.  We take the live time to be 988 days, and use the effective detector mass from~\cite{Aartsen:2013jdh}, which takes into account how the sensitivity drops at low energy, and flattens at $\sim$\,0.4\,Gton at high energy.  We take the neutrino cross section from~\cite{Gandhi:1998ri}, and assume 10\% detector energy resolution.  We do not show results for Models A and D because they do not show appreciable differences in this energy range.  

Given the low statistics, all spectra in Fig.~\ref{fig:binspec} describe the data points reasonably well.  However, with more exposure, IceCube should be able to distinguish these $\nu$SI cases from a power-law spectrum.  Similarly, the flux spectra shown in Fig.~\ref{fig:astro2} and Fig.~\ref{fig:astro3} might also be distinguishable in the future.  Other than associating the gap with a resonance dip, it is also interesting to ask if the observed events near $\sim1$\,PeV energies can be caused by a bump-like feature as shown in Fig.~\ref{fig:astro2}.  This hypothesis can be tested by fitting a precise cascade spectrum with $\nu$SI parameters such as cross section and mediator mass.  It is interesting to note that the required parameter are not too far away from laboratory constraints~\cite{Ioka:2014kca}.

We anticipate future IceCube analyses can extend to lower energies with better background rejection.  In Fig.~\ref{fig:ednde}, we show the cascade event spectrum for a 1-Gton detector with the 988 days and the same detector energy resolution as above, but with 10 bins per decade to match the 10\% energy resolution.  We consider perfect detector efficiency.  We see that Models B and C cause distinct spectral features that are detectable.  However, it is clearly challenging to distinguish Models A and D. 

The flavor phenomenology described above varies mildly with the uncertainty in mixing parameters.  In principle, it can be self-consistently incorporated into three-flavor Boltzmann equations by generalizing Eq.~(\ref{eq:boltzmann}).  Another class of scenarios we would like to mention is for the coupling to be in the mass basis instead of the flavor basis.  The spectral evolution in that case would also depend on the mass difference as well as mass hierarchy.  In this work we only wish to highlight the phenomenology of spectral distortion by $\nu$SI.  We defer the comprehensive analysis for studies of concrete models, which may also find interesting flavor effects. 

\subsection{Further discussion}

The results above demonstrate that $\nu$SI could appreciably affect the spectrum detected by IceCube, perhaps in ways that could explain some of its features.  However, it is too soon to make definite statements.  The most obvious point is that the IceCube data is presently sparse but will soon improve in both statistics and the types of searches~(cascades and tracks, diffuse and point sources, etc.).  Combined with multi-messenger studies, this will help identify the origin of the events and thus information about their emitted spectrum.  Once there is more data, more detailed calculations will be warranted.  Those could explore a wider range of theoretical possibilities for the $\nu$SI scenarios.

As described above, we assume that just one species (flavor or mass eigenstate) of neutrinos and antineutrinos experiences $\nu$SI.  This is because the laboratory limits are strong for $\nu_e$ and $\nu_\mu$, but weak for $\nu_\tau$.  Thus our calculations are nominally for $\nu_\tau + \bar{\nu}_\tau$.  However, the situation is more complex.  Because of the vast distances, astrophysical neutrinos propagate as incoherent mass eigenstates, and all of the mass eigenstates have an appreciable $\nu_\tau$ fraction.  Whether the primarily $\nu$SI couplings are to flavor or mass eigenstates is model-dependent.   The effects of $\nu$SI that we illustrate for one species will be diluted by the lack of $\nu$SI for the other species, but the details are model dependent.  A closer look at how the laboratory and astrophysical studies together constrain the different flavor or mass eigenstates is needed.  Flavor ratios for astrophysical neutrinos may be an especially important test.

If the neutrino masses are not degenerate, then resonances could occur at different energies, which would lead to more complicated spectra or possibly even cancelations between dips and bumps.  In the case that the lightest neutrino is relativistic today, there would be non-negligible thermal broadening.  There could also be model-dependent details that complicate the discussion, including by having more than one mediator, by coupling to dark matter, or by having more general final states.  We took both final states to be active neutrinos.  If only one is, then the absorption dip and spectral cutoff would be unaffected but the cascade bump would be reduced.  If neither is, then only the absorption dip or spectral cutoff is observable.

\section{Conclusions and outlook}
\label{sec:conclusion}

Neutrinos may still hold surprises, and $\nu$SI are among the possibilities.  Their effects can be probed directly through neutrino-neutrino scattering --- provided that we have detected neutrinos from astrophysical sources traveling through the C$\nu$B.  Until recently, this was only possible with the SN 1987A data~\cite{Kolb:1987qy}.  The detection of high-energy neutrinos by IceCube has opened a new frontier in neutrino astronomy, which provides new opportunities for probing $\nu$SI.  Because the IceCube sources appear to be extragalactic, the column density of neutrino targets is much greater than for SN 1987A; because the energies are much larger, a wider range of $\nu$SI parameters can be probed; and, because the observed flux is diffuse, that averages out the peculiarities of individual sources.

The observed IceCube spectrum contains interesting features, which include a gap at moderate energies, a possible excess near 1 PeV, and a cutoff at slightly higher energies~\cite{Aartsen:2013jdh, Aartsen:2013bka, Aartsen:2014gkd}.  Given the current statistics, these features are consistent with standard model expectations with simple astrophysical assumptions~\cite{Laha:2013lka, Chen:2013dza, Anchordoqui:2013dnh}.  It is, however, interesting to consider exotic explanations such as $\nu$SI, pseudo-Dirac neutrinos~\cite{Esmaili:2012ac}, or Lorentz-invariance violation~\cite{Gorham:2012qs, Anchordoqui:2014hua}. 

We perform the first study of $\nu$SI in the context of the detected IceCube spectrum and its features.  Using a phenomenological approach for the interactions, we show that IceCube is sensitive to an interesting range of $\nu$SI parameters that evades the most robust of the laboratory limits and is more sensitive than other astrophysical or cosmological techniques.  We provide an improved calculation using the propagation equation, the first for high-energy neutrinos to take into account $\nu$SI through attenuation, regeneration, and multiple scattering.  Solving the propagation equation numerically, we show $\nu$SI could generate spectral distortions such as a dip, bump, or cutoff large enough to mimic the features seen in the IceCube spectrum.  Although $\nu$SI might be able to explain some features of the observed data, it is too soon to draw such conclusions.  We expect the IceCube spectrum will become more precise in the near future by improved statistics and analysis.  With that, more detailed phenomenological studies and associated model-building will be possible.

An expected --- but not yet observed --- source of high-energy astrophysical neutrinos is produced through the energy losses of ultra-high-energy cosmic rays propagating through the CMB~\cite{Greisen:1966jv, Zatsepin:1966jv}.  Once these cosmogenic neutrinos~\cite{Beresinsky:1969qj} are observed, it will be possible to test $\nu$SI using calculations similar to those presented here.  Although the cosmogenic neutrino spectrum is not a simple power law, its shape is reasonably well predicted.  For an energy of $\sim 10^{10} {\rm\ GeV}$, a resonance with the C$\nu$B would probe mediator masses near $M \sim 10^3$ MeV, which are not well constrained (see Fig.~\ref{fig:constraint}).  We do not show the curves for optical depth $\tau$ for this case; their shape is similar to that for PeV neutrinos, but displaced to larger mediator masses and couplings (for a heavy mediator, the sensitivity is $g/M\sim0.4/\left(10^{3}\,{\rm MeV} \right)$).  Could $\nu$SI explain the non-observation of cosmogenic neutrinos?  While pushing their spectrum to lower energies could be consistent with IceCube data, the required coupling is relatively large, $g\sim 1$.

Our calculations are for a diffuse flux, which is consistent with IceCube data.  If point sources are observed, the effects of deflection and delay should be noted (these are irrelevant for the diffuse flux).  The Lorentz factor $\gamma$ of the center of momentum frame is $\sim 10^{8}$ for 1 PeV neutrinos scattering on neutrinos of mass 0.1 eV.  For one scattering, the deflection is $\Delta \theta \sim10^{-8} \, (10^{8}/\gamma)$, which is tiny, and the time delay is $\Delta t \sim 10 {\rm\ s} \,(10^8/\gamma)^2$, which might not be negligible in some cases.  These effects would be increased by multiple scattering, ultimately washing out transient and point sources into a steady diffuse flux.  For reasonable couplings, these effects are not relevant for the PeV neutrinos.

For low-energy neutrinos from a nearby supernova, these effects could be much more important.  The delay is smaller by $\sim 10^6$ due to the closer distance but larger by $\sim 10^8$ due to the change in $\gamma^2$, making $\Delta t \sim 10^3 {\rm\ s}$.  KT87~\cite{Kolb:1987qy} defined their constraint by changes in the energy due to energy loss, which requires assumptions about the total energy in neutrinos and the energy spectrum.  The same constraint can be obtained by the simpler time delay argument, which only requires an assumption about the total energy in neutrinos.

The IceCube neutrino telescope has opened a new age in neutrino astronomy, as well as providing a way to directly test $\nu$SI.  Complementary constraints should also be developed for neutrinos in the early universe and core-collapse supernovae.  In those settings, even weak-scale neutrino-neutrino collisions and mixing from the self-induced potential are important.  The rapid advance of precision cosmology and perhaps a lucky detection of a Milky Way supernova might reveal more secrets about neutrinos.

\medskip
{\bf Note added:}  As this paper was being completed, we learned of an independent study by Ioka and Murase~\cite{Ioka:2014kca}, which was submitted to arXiv simultaneously. 

\section*{Acknowledgments}
We thank the participants of the CCAPP workshop on ``Cosmic Messages in Ghostly Bottles: Astrophysical Neutrino Sources and Identification,'' especially Markus Ahlers, Kfir Blum, Kohta Murase, and Mary Hall Reno, for helpful comments.  We also thank Yasaman Farzan, Shirley Li, Shmuel Nussinov, and Sergio Palomares-Ruiz, for helpful comments.  We are especially gratefully to Ranjan Laha for participation in an early phase of this work and for many helpful discussions since.  K.C.Y.N. and J.F.B. were supported by NSF Grant PHY-1101216 to J.F.B.

\bibliographystyle{h-physrev}
\bibliography{references}

\end{document}